\documentclass[a4paper,12pt]{article}
\usepackage{a4wide}
\usepackage[utf8]{inputenc}
\usepackage{graphicx}
\usepackage{amsmath}
\usepackage{amssymb}
\usepackage{array}
\usepackage{hyperref}
\usepackage{tabularx}
\usepackage{cite}

\begin{document}

\begin{titlepage}
\setcounter{page}{0}
\begin{flushright}
May 2020 \\
\end{flushright}
\vfill
\begin{center}

{\large\bf Chiral Perturbation Theory at NNNLO}
\\[3cm]
{\bf Nils Hermansson-Truedsson}
\\[5mm]
	{Universit\"{a}t Bern, Albert Einstein Center for Fundamental Physics, Institute for Theoretical Physics, Sidlerstrasse 5, 3012 Bern, Switzerland, \\nils@itp.unibe.ch}	
\\[5mm]
	{Prepared for submission to \emph{Symmetry}, for the special issue \emph{Effective Field Theories - Chiral Perturbation Theory and Non-relativistic QFT}}
\vfill

{\large\textbf{Abstract}}
\end{center}
	Chiral perturbation theory is a much successful effective field theory of quantum chromodynamics at low energies. The effective Lagrangian is constructed systematically order by order in powers of the momentum $p^2$, and until now the leading order (LO), next-to leading order (NLO), next-to-next-to leading order (NNLO) and next-to-next-to-next-to leading order (NNNLO) have been studied. In the following review we consider the construction of the Lagrangian and in particular focus on the NNNLO case. We in addition review and discuss the pion mass and decay constant at the same order, which are fundamental quantities to study for chiral perturbation theory. Due to the large number of terms in the Lagrangian and hence low energy constants arising at NNNLO, some remarks are made about the predictivity of this effective field theory.

\end{titlepage}

\section{Introduction}
At low energies the strong force of Nature is formulated as Quantum chromodynamics (QCD), a quantum field theory whose degreees of freedom are the fundamental quarks and gluons. These particles are not observable, but confined in composite particles known as hadrons. The hadrons are divided into classes depending on their quark content, and in the following we shall be concerned with the mesons, i.e.~bound states of a quark and an antiquark. The mesons of relevance here are the pions, the kaons and the eta meson. At low energies QCD is non-perturbative, i.e.~it is not possible to calculate observables systematically order by order in the strong coupling constant $\alpha _{s}$. However, chiral perturbation theory (ChPT)~\cite{Weinberg:1978kz,Gasser:1983yg,Gasser:1984gg} is an effective field theory (EFT) of QCD valid at low energies. The systematic expansion is done in powers of the momentum $p^2$ and is expected to hold up to energy scales on the order of $1$ GeV. Systematic is here a keyword, since one may improve the precision in a calculation by increasing the number of terms in the expansion. However, each order comes with increased computational complexity and for a prediction requires the knowledge of certain low energy constants (LECs). In the following we shall review the current status of this expansion. 

The construction of ChPT is based on the approximate chiral symmetry of QCD,
\begin{align}
	SU(N_{f})_{L}\times SU(N_{f})_{R} \, ,
\end{align}
which holds in the massless quark limit for $N_{f}$ quarks. The physically relevant cases are $N_{f}=2$ when one considers only the up and down quarks, and $N_{f}=3$ when also the strange quark is included. In particular, one exploits the spontaneous breaking of the chiral symmetry from a non-vanishing quark-antiquark condensate $\langle \bar{q}q \rangle \neq 0$, i.e. 
\begin{align}
	SU(N_{f})_{L}\times SU(N_{f})_{R} \longrightarrow SU(N_{f})_{V}\, ,
\end{align}
and builds an effective Lagrangian $\mathcal{L}_{\chi}$ from this. The Lagrangian is Lorentz invariant and also satisfies the discrete symmetries of the QCD Lagrangian. The operators of the Lagrangian are constructed from building blocks containing the $N_{f}^2-1$ lightest pseudoscalar mesons and external source fields, as will be explained in the next section. For the two-flavour case the mesonic degrees of freedom are the pions $\pi^{\pm}$ and $\pi ^{0}$, whereas for three flavours one has also the kaons and the eta, namely $K^{\pm}, \, K^{0}, \, \bar{K}^{0}$ and $\eta$. 

The Lagrangian is written down order by order in powers of $p^2$ and can therefore be written as
\begin{align}
	\mathcal{L}_{\chi} = \sum _{n=1}^{\infty} \mathcal{L}_{2n} \, ,
\end{align}
where each Lagrangian $\mathcal{L}_{2n}$ has the form
\begin{align}
	\mathcal{L}_{2n} = \sum _{i=1} ^{N_{2n}} c_{i}^{(2n)}\mathcal{O}_{i}^{(2n)} \, .
\end{align}
The $N_{2n}$ so-called monomials have here been denoted $\mathcal{O}_{i}^{(2n)}$ and are of chiral order $p^{2n}$. The coefficients $c_{i}^{(2n)}$ are the LECs referred to earlier. As can be seen, each order introduces new LECs and in order to make numerical predictions one needs to know these coefficients. This is discussed further in the next section. 

The leading order (LO) and next-to leading order (NLO) Lagrangians have been known for a long time~\cite{Weinberg:1978kz,Gasser:1983yg,Gasser:1984gg}. The next-to-next-to leading order (NNLO) Lagrangian was derived some time later~\cite{PhysRevD.53.315,Bijnens:1999sh}. It was, however, not until recently that the next-to-next-to-next-to leading order (NNNLO) Lagrangian was derived~\cite{Bijnens:2018lez} and calculations at this order were started~\cite{Bijnens:2017wba}. These references only consider the non-anomalous sector. The anomalous NLO case was considered in Refs.~\cite{Wess:1971yu,Witten:1983tw}, and NNLO in Refs.~\cite{Ebertshauser:2001nj,Bijnens:2001bb}. The anomalous NNNLO Lagrangian has so far not been considered.

In the following we shall review the current status of ChPT at NNNLO, but due to the scarce literature this review will focus on the construction of the Lagrangian and the pion mass and decay constant at NNNLO.

 
\section{Constructing the Effective Lagrangian}
In constructing the effective chiral Lagrangian at an order $p^{2n}$, one has to start from the spontaneous breakdown of the chiral symmetry, i.e.~$G\rightarrow H$ where $G = SU(N_{f})_{L}\times SU(N_{f})_{R}$ and $H = SU(N_{f})_{V}$. In this symmetry breaking $N_{f}^2-1$ generators of $G$ are broken, and by virtue of Goldstone's theorem there are $N_{f}^2-1$ associated Goldstone bosons. For ChPT these are the $N_{f}^2-1$ lightest pseudoscalar mesons. They are associated to the elements in the coset space $G/H$~\cite{PhysRev.177.2247}. A detailed discussion of this correspondence will here be left out as it is not of central importance. The most important observation is, however, that there is choice in how to represent the Goldstone bosons and one can write down $\mathcal{L}_{2n}$ in any such basis. The physical predictions are of course the same. 

Each basis is defined in terms of building blocks that contain the meson fields as well as external fields, and when constructing the Lagrangian one should combine these buildings blocks to construct the basis monomials\footnote{For notational convenience, we hereby refer to the $\mathcal{O}_{i}^{(2n)}$ as monomials rather than operators. The reason is that operators constructed from the chiral symmetry not necessarily satisfy the other symmetries and therefore need to be linearly added in order to do so.} invariant under all imposed symmetries. For QCD this means chiral symmetry, Lorentz invariance as well as hermitian conjugation ($\textrm{H.C.}$), parity ($\mathcal{P}$) and charge conjugation ($\mathcal{C}$). The key in constructing the Lagrangian is thus to analyse how the building blocks transform under $G$, and bearing in mind their respective scalings with $p^2$.

Let us next consider how to construct the effective Lagrangian in some detail. Denote a general group element in $G$ as $(g_{L},g_{R})$ and in addition let $h\in H$. The meson fields are included in the building blocks $u$ and $U = u^2$ transforming under $G$ as
\begin{align}\label{eq:blocks1}
	\mathcal{O}\big(p^0\big): \; \;  & u\longrightarrow g_{R} u \, h ^{\dagger} = h \, u g_{L}^{\dagger} \, , \nonumber \\
	 \mathcal{O}\big( p^0\big):  \; \;  & U\longrightarrow g_{R} U \, g_{L}^{\dagger} \, .
\end{align}
It has here been indicated that both scale as $p^0$ in the chiral counting. As an example of how the meson fields enter, the matrix $U$ can be written in an exponential parametrisation according to $U = \exp \{ i T^a\phi ^a /F\} $, where $T^a$ are $SU(N_{f})$ generators and the meson fields are denoted $\phi ^{a}$. One may therefore write
\begin{align}
N_{f}=3 \, : \; \; \; T^a \phi ^a =
\left( 
\begin{array}{ccc}
\pi ^{0}+\frac{1}{\sqrt{3}}\eta & \sqrt{2} \, \pi ^{+} &  \sqrt{2}\, K^{ +}\\
 \sqrt{2} \, \pi ^{-} & -\pi ^{0} +\frac{1}{\sqrt{3}}\eta &  \sqrt{2}\,  K^{ 0}\\
\sqrt{2}\, K^{ -} & \sqrt{2}\, \overline{K}^{ 0} & -\frac{2}{\sqrt{3}}\eta
\end{array}
\right) \, , 
\end{align}
and
\begin{align}
N_{f}=2 \, : \; \; \; T^a \phi ^a =
\left( 
\begin{array}{cc}
\pi ^{0} & \sqrt{2} \, \pi ^{+} \\
 \sqrt{2} \, \pi ^{-} & -\pi ^{0}
\end{array}
\right) \, . 
\end{align}

The external fields used to construct building blocks are scalar ($s$), pseudoscalar ($p$), vector ($v_{\mu}$) and axial vector ($a_{\mu}$), respectively.  
These are combined into
\begin{align}\label{eq:blocks2}
	\mathcal{O}\big(p^2\big): \; \; & \chi = 2B (s+ip)  \longrightarrow  g_{R}\chi g_{L}^{\dagger} \, , \nonumber \\
	\mathcal{O}\big(p\big): \; \; & \ell _{\mu} = v_{\mu}-a_{\mu} \longrightarrow g_{L} \ell _{\mu} g_{L}^{\dagger} -i \,  \partial _{\mu} g_{L} g_{L}^{\dagger} \, , \nonumber \\
	\mathcal{O}\big(p\big): \; \; & r _{\mu} = v_{\mu}+a_{\mu}  \longrightarrow g_{R} r _{\mu} g_{R}^{\dagger} -i \, \partial _{\mu} g_{R} g_{R}^{\dagger} \, , 
\end{align}
where $B = -\langle \bar{q}q\rangle /\Big( N_{f}F^{2} \Big) $ is a constant depending on the pion decay constant $F$. One may further define field strength tensors via $\ell _{\mu}$ and $r_{\mu}$, but these depend on the basis. For the moment, let us consider the canonical choice of Refs.~\cite{Gasser:1983yg,Gasser:1984gg} for the construction of the NLO Lagrangian. This choice will in the following be called the $U$ basis. There one has 
\begin{align}\label{eq:Ufsblocks}
	\mathcal{O}\big(p^2\big): \; \; & F_{L}^{\mu \nu}  \longrightarrow  \partial ^{\mu} \ell ^{\nu } - \partial ^{\nu} \ell ^{\mu} -i \left[ \ell ^{\mu},\ell ^{\nu} \right] \longrightarrow g_{L}F^{\mu \nu}_{L}g_{L}^{\dagger} \, , \nonumber \\
	\mathcal{O}\big(p^2\big): \; \; & F_{R}^{\mu \nu}  \longrightarrow  \partial ^{\mu} r ^{\nu } - \partial ^{\nu} r ^{\mu} -i \left[ r ^{\mu}, r ^{\nu} \right] \longrightarrow g_{R}F^{\mu \nu}_{R}g_{R}^{\dagger} \, .
\end{align}
In addition to the above field strength tensors one also introduces the covariant derivative acting on an operator $O$ constructed from the building blocks in the $U$ basis according to
\begin{align}
	\label{eq:Ucovder}
D_{\mu} O =
\left\{
\begin{array}{ll}
 \partial _{\mu}O -ir_{\mu}O +iO \ell _{\mu},  & O\longrightarrow g_{R}\, O\, g_{L}^{\dagger}  
\, ,  \\ \\ 
\partial _{\mu}O  -i \ell _{\mu}O+iOr_{\mu},  & O\longrightarrow g_{L}\, O\, g_{R}^{\dagger}  
\, ,  \\ \\
\partial _{\mu}O -ir_{\mu}O +i O r _{\mu},  & O\longrightarrow g_{R}\, O\, g_{R}^{\dagger}  
\, ,  \\ \\
\partial _{\mu}O -i \ell_{\mu}O +i O\ell _{\mu},  & O\longrightarrow g_{L}\, O\, g_{L}^{\dagger}  
\, .  \\
\end{array}
\right.
\end{align}
	The covariant derivative has chiral order $p$ and as can be seen transforms differently under $G$ depending on the operator $O$. As will be discussed further below, this requires special consideration for the generation of chirally invariant monomials including $D_{\mu}$, which in particular is something that one may want to avoid for higher order Lagrangians.  

Knowing the building blocks in eqs.~(\ref{eq:blocks1})-(\ref{eq:Ufsblocks}), it is now possible to derive the various chiral Lagrangians. The steps in doing so are
\begin{enumerate}
	\item Write down all possible operators satisfying chiral symmetry and Lorentz invariance, 
	\item Create linear combinations of these operators into monomials such that $\textrm{H.C}$, $\mathcal{C}$ and $\mathcal{P}$ are satisfied,
	\item Eliminate equivalent monomials which are related by certain operator identities.
\end{enumerate}
The first point is done by combining the building blocks for a given chiral $p^{2n}$ in all possible ways. In practice this means taking flavour space traces of, and products of traces, of products of building blocks. The resulting structures do not necessarily satisfy the discrete symmetries, and one must in general create linear combinations of the generated operators, i.e.~the second step is required, and monomials satisfying the discrete symmetries are obtained. Transformation properties of the building blocks can be found in many places, e.g.~in Ref.~\cite{Bijnens:2018lez}, and are therefore left out here. The final remark here is particularly important. Following the first two steps one creates an initial number, $N_{2n}^{0}$, say, of monomials. However, these monomials can be related via a set of relations, which for ChPT up to NNNLO are
\begin{enumerate}
	\item LO equations of motion, or, equivalently, field redefinitions~\cite{Scherer:1994wi,Bijnens:2001bb},
	\item Integration by parts (IBP) identities,
	\item The Bianchi identity,
	\item Specific operator identites such as commutation relations of covariant derivatives,
	\item $N_{f}$--specific relations from the Cayley-Hamilton theorem.
\end{enumerate}
Beyond NNNLO, one will also have to include other relations. An example is the so-called Schouten identity~\cite{Schouten:1938}, stating that it is impossible to create a completely antisymmetric tensor with more indices than the number of dimensions\footnote{At NNNLO one can in principle have up to eight Lorentz indices. However, due to Lorentz invariance, there can be no more than four indpendent indices. As a consequence, in four dimensions the Schouten identity only gives monomial relations starting from order $p^{10}$.}. The relations will for simplicity not be discussed in detail here, but are reviewed thoroughly in Ref.~\cite{Bijnens:2018lez}. Each of them is linear, and together the linear equations form a system of linear equations. This can be written
\begin{align}\label{eq:soe}
	0 = \mathcal{R} \vec{\mathcal{O}}^{(2n)} \, ,
\end{align}
where $\mathcal{R}$ is a matrix of coefficients which when acting on the vector of $N_{2n}^{0}$ monomials $ \mathcal{O}^{(2n)}_{i}$ gives $N_{\textrm{rel}}$ relations. In other words, the matrix has size $N_{\textrm{rel}}\times N_{2n}^0$. Each relation gives the option to replace one monomial by a linear combination of the others, i.e.~$N_{2n}\leq N_{2n}^0$. The specific relations are, however, not necessarily independent. Therefore, one has to find a set of linearly independent relations and for this purpose one can use Gaussian elimination in~(\ref{eq:soe}). It follows that
\begin{align}
	N_{2n} = N_{2n}^0 -\textrm{rank} (\mathcal{R}) \, ,
\end{align}
and from here a minimal basis is obtained, i.e.~the number of LECs and hence monomials has been minimised. In case this minimisation is not performed, the number of redundant operators may be very large. For instance, at NNNLO the basis for a general $N_{f}$ shrinks from $N_{2n}^{0}\sim 10000$ to $N_{2n}\sim 2000$ (see the next section). However, the LECs in such a basis of course combine such that only linear combinations corresponding to the LECs in the minimal basis appear.  

An important point regarding the minimality of the operator basis must also be made. It is in general hard to know that a minimal basis has been obtained and the above method works as long as all relations are known. It also provides a computationally straightforward way of attacking the problem. One method guaranteed to find a minimal basis is based on constructing all possible Green's functions in order to determine LECs. This is discussed in Ref.~\cite{Ruiz-Femenia:2015mia} for the NNLO Lagrangian. A final point to make here is that it also is possible to use a so-called Hilbert series approach, see Ref.~\cite{Henning:2017fpj}, but this will not be discussed further in this review as it has not been applied at NNNLO. Naturally, all approaches must give the same number.

\subsection{The Lagrangians of Chiral Perturbation Theory}

Following the above prescription with writing down a linear system of relations between monomials, one can start deriving the chiral Lagrangians. At LO there are only three possible monomials to write down. In the $U$ basis one readily finds
\begin{align}\label{eq:UbasisL2}
	\mathcal{L}_{2} = \frac{F^{\, 2}}{4} \Big\langle \left( D_{\mu}U\right) ^{\dagger} D^{\mu} U \Big\rangle + \frac{F^{\, 2}}{4} \Big\langle \chi ^{\dagger} U + \chi U^{\, \dagger} \Big\rangle \, ,
\end{align}
where $\langle \ldots \rangle$ denotes the trace in flavour space. The third monomial left out here is $\Big\langle U ^{\dagger}D^{2} U + (D^2 U)^{\dagger} U\Big\rangle$, and was eliminated by IBP.

The same procedure can be used at every order in the chiral expansion, but two interesting features appear at NLO. First of all, it becomes possible to write down so-called contact terms, i.e.~terms without dependence on the meson fields. These terms are not physical but still appear at every order. Examples of contact terms at order $p^8$ are
\begin{align}
	& \big\langle \chi \chi ^{\dagger}\rangle \langle \chi ^{\dagger} \chi \big\rangle \, ,
	\nonumber \\
        & i \big\langle D_{\mu}\chi D_{\nu} \chi ^{\dagger} F_{R}^{\mu \nu} +  D_{\mu} \chi ^{\dagger} D_{\nu} \chi  F_{L}^{\mu \nu}\big\rangle \, .
\end{align}
In the first of these one sees that also products of traces can contribute to the Lagrangian. The second example is particularly interesting as it highlights how linear combinations of separately chirally invariant operators must be linearly combined to satisfy $\textrm{H.C.}$, $\mathcal{C}$ and $\mathcal{P}$. This can also be seen in~(\ref{eq:UbasisL2}).

In addition to the existence of counterterms, the size of the minimal basis depends on $N_{f}$. The reason comes from the Cayley-Hamilton theorem which states that the $N_{f}\times N_{f}$ matrices satisfy their own characteristic equation. The products of building blocks in the monomials therefore also satisfy this, and one obtains different numbers of relations for $N_{f}=2$ and $N_{f}=3$, as mentioned in the previous section.

The $U$ basis is not necessarily the most suitable choice for higher order Lagrangians. The reason is that the building blocks all transform differently, i.e.~it is not possible to generate all chirally and Lorentz invariant operators by creating a list of all possible permutations of building blocks. An laternative to the $U$ basis is the $u$ basis, where the building blocks are
\begin{align}
\label{eq:ubuildingblocks}
	& \mathcal{O}(p): \; \;  u_{\mu} = i\Big[ u^{\dagger} (\partial _{\mu}-ir_{\mu})u-u(\partial _{\mu}-i\ell _{\mu} )u^{\dagger}\Big] \longrightarrow h u_{\mu} h^{\dagger}
\, ,
	\nonumber \\
	& \mathcal{O}(p^2): \; \; \chi _{\pm} = u^{\dagger}\chi u^{\dagger}\pm u\chi ^{\dagger} u \longrightarrow h \chi _{\pm} h^{\dagger}
\, , 
	\nonumber\\
	& \mathcal{O}(p^2): \; \; f_{\pm}^{ \mu \nu} = u F_{L}^{\mu \nu }u^{\dagger} \pm u^{\dagger}F_{R}^{\mu \nu}u \longrightarrow h f_{\pm} ^{\mu \nu} h^{\dagger} 
 \, .
\end{align}
As can be seen, each of the building blocks transforms in the same way. As a consequence, taking the flavour trace of any combination of building blocks will automatically be chirally invariant. One also defines a covariant derivative $\nabla _{\mu}$ according to
\begin{align}
	& \nabla _{\mu} O  = \partial _{\mu} O + \big[\Gamma _{\mu}, O \big] \, , 
	\nonumber \\
	& \Gamma _{\mu} = \frac{1}{2}\Big[ u^{\dagger}\big( \partial _{\mu}-ir_{\mu}\big) u + u \big( \partial _{\mu}-i \ell _{\mu}\big) u^{\dagger}  \Big] \, ,
\end{align}
where $O$ is any operator transforming as $O\rightarrow h O h^{\dagger}$. Evidently, this covariant derivative has a much simpler form than $D_{\mu}$ in~(\ref{eq:Ucovder}).

The LO Lagrangian in the $u$ basis is given by
\begin{align}
	\mathcal{L}_{2} = \frac{F^{\, 2}}{4} \Big\langle  u_{\mu} u^{\mu} + \chi _{+} \Big\rangle 
\, .
\end{align}
Note that there are no covariant derivatives appearing here, since no Lorentz invariant terms of order $p^2$ including $\nabla _{\mu}$ can be written down. Also, no linear combinations of operators are needed for the discrete symmetries. 

The number of terms and hence LECs in the minimal basis for each order up to NNNLO is presented in Table~\ref{table:noterms}. The number increases rapidly, but the contact terms still remain few. An interesting question is how many terms appear when one removes external fields. There are three cases, the first where only $\chi _{\pm}$ are kept, the second where only $f_{\pm}^{\mu \nu}$ remain and the third where no external fields appear. The numbers are presented in Table~\ref{table:nnnlonumbers}. Comparing this to Table~\ref{table:noterms}, one sees that the number of terms reduces drastically with the exclusion of external fields. The case with only $\chi _{\pm}$ is interesting for the calculation of the masses and decay constants in the isospin symmetric case, as will be explained in the next section. An example of the NNNLO Lagrangian is given in the supplementary material of Ref.~\cite{Bijnens:2018lez}.

\begin{table}[t]
\begin{center}  
\renewcommand{\arraystretch}{1.2}
\caption{The number of monomials in the chiral Lagrangians up to NNNLO. Here, all external fields are included and the number of contact terms have also been indicated.}
\begin{tabular}{ccc|cc|cc}
   &\multicolumn{2}{c}{$N_{f}$} &\multicolumn{2}{c}{$N_{f}=3$} &\multicolumn{2}{c}{$N_{f}=2$} \\ \hline 
      & Total & Contact & Total & Contact & Total & Contact\\ \hline
$p^2$ & 2     & 0       & 2     &  0      & 2     & 0 \\
$p^4$ & 13    & 2       & 12    &  2      & 10    & 3 \\
$p^6$ & 115   & 3       & 94    &  4      & 56    & 4 \\
$p^8$ & 1862  & 22      & 1254  &  21     & 475   & 23\\ \hline
    \end{tabular}
   \renewcommand{\arraystretch}{1}
	\label{table:noterms}
  \end{center}
\end{table}

\begin{table}[t]
\begin{center}  
\renewcommand{\arraystretch}{1.2}
\caption{The number of monomials in the chiral Lagrangian at NNNLO. Here, only the external fields indicated in the left column have been included. }
	\begin{tabular}{ccc|cc|cc}
   &\multicolumn{2}{c}{$N_{f}$} &\multicolumn{2}{c}{$N_{f}=3$} &\multicolumn{2}{c}{$N_{f}=2$} \\ \hline 
	& Total & Contact & Contact & Total & Total & Contact\\ \hline
$\chi_{\pm}$ &   538   &  3      & 328     & 4       & 122     & 6  \\
	$f_{\pm}^{\mu \nu}$ &  963  & 15       &  591   & 13      &  238   & 11  \\
None &  135  &    0    &  56   &  0     & 16    & 0  \\
    \end{tabular}
   \renewcommand{\arraystretch}{1}
	\label{table:nnnlonumbers}
  \end{center}
\end{table}

\subsection{The predictivity of chiral perturbation theory}
The predictivity of ChPT relies on the knowledge of the LECs appearing in the Lagrangian. Disregarding contact terms in the Lagrangian, at NLO the LECs are the $L_{i}$ and $\ell _{i}$ for the three-flavour and two-flavour theories, respectively,~\cite{Gasser:1983yg,Gasser:1984gg}, and are of course divergent. Through renormalisation one obtains the finite and scale dependent parameters $L_{i}^{r}(\mu)$ and $\ell _{i}^{r}(\mu)$, where $\mu$ is the renormalisation scale, which are the values needed for predictions starting at NLO. It is possible obtain values for them e.g.~from experimental data by making fits, from large $N_{c}$ or vector meson dominance approaches~\cite{Pich:2018ltt,Bijnens:2014lea,Bijnens:1997vq,Ecker:1989yg,ECKER1989311}.  

The renormalisation of the NNLO LECs was performed in Ref.~\cite{Bijnens:1999hw}. The LECs there are denoted $C_{i}$ for $N_{f}=3$ and $c_{i}$ for $N_{f}=2$, and the renormalised values are not known at all as well as the NLO LECs. However, certain combinations of the LECs can be constrained, such as in Ref.~\cite{Bijnens:1997vq} where combinations in $\pi\pi$ scattering were estimated with resonance saturation. 

At NNNLO, the only renormalisation so far performed is for the two-flavour case and can be found in Ref.~\cite{Bijnens:2017wba}. As will be explained below, two combinations of NNNLO LECs show up for the pion mass and decay constant. By using the results from Refs.~\cite{Gasser:1983yg,Bijnens:1999hw} and that physical quantities are finite, the two NNNLO combinations could be renormalised in Ref.~\cite{Bijnens:2017wba}. It would of course be interesting to perform the full renormalisation at NNNLO. However, a natural question is how feasible it would be to actually obtain numerical values for the renormalised LECs at NNNLO, especially seeing the status for the NNLO LECs.    

An additional important source for LECs not discussed above comes from lattice gauge theory. On the lattice, quark masses can be varied and if one considers the ChPT predictions as functions of these parameters one can obtain information not available from experiments. For a review of the current status of obtaining LECs from the lattice, see Ref.~\cite{Aoki:2019cca}. 

\section{The Pion Mass and Decay Constant}
Two quantities of fundamental interest in ChPT are the pseudoscalar masses and decay constants. These quantitites can be calculated order by order in the chiral expansion. At NLO this requires a one-loop computation, at NNLO a two-loop one and at NNNLO it means three-loop diagrams also have to be included. For $SU(3)$, several mass scales occur which make the loop integral calculation substantially harder. The NLO quantities were calculated already in Ref.~\cite{Gasser:1983yg,Gasser:1984gg} and at NNLO in Refs.~\cite{Burgi:1996qi,Bijnens:1997vq,Bijnens:1995yn,Amoros:1999dp,Ananthanarayan:2017yhz,Ananthanarayan:2018irl}. The pion mass and decay constant were calculated at NNNLO for isospin symmetric two-flavour ChPT in Ref.~\cite{Bijnens:2017wba}, but are hitherto unknown for $SU(3)$ due to the lack of knowledge of the appearing three-loop integrals. Below, we discuss the NNNLO results.

\begin{figure}[htbp]
\begin{minipage}{0.48\textwidth}
\includegraphics[width=0.99\textwidth]{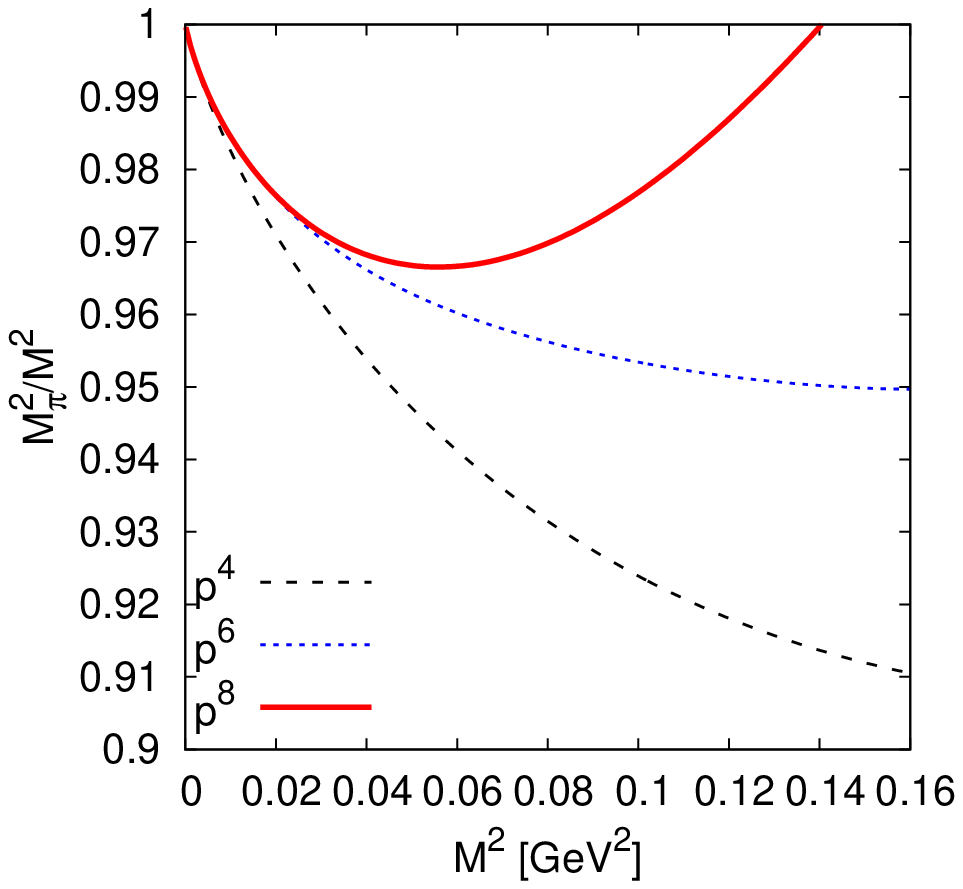}\\[-3mm]
\centerline{(a)}
\end{minipage}
\begin{minipage}{0.48\textwidth}
\includegraphics[width=0.99\textwidth]{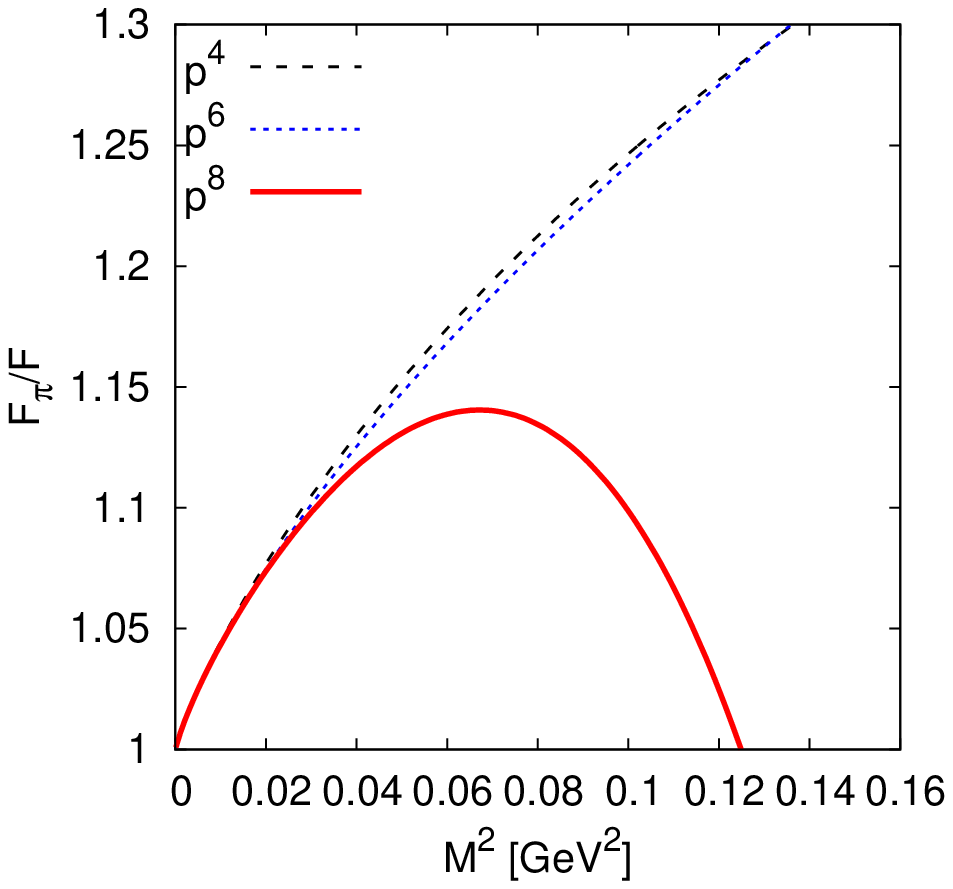}\\[-3mm]
\centerline{(b)}
\end{minipage}
	\caption{(a) The pion mass $M_{\pi}^2/M^2$, and (b) the decay constant $F_{\pi}/F$ at NLO, NNLO and NNNLO, respectively. By definition, the LO is identically unity. This plot is taken from Ref.~\cite{Bijnens:1997vq}.}
\label{fig:mqdep}
\end{figure}

The pion mass and decay constant are expanded order by order according to
\begin{align}
	& M_{\pi}^2 = M^2\big( 1+M_{4}^2+M_{6}^2+M_{8}^2\big) \, ,
        \nonumber \\
	& F_{\pi} = F\big( 1+F_{4}+F_{6}+F_{8}\big) \, ,
\end{align}
where the physical pion mass $M_{\pi}^2$ is defined as the pole of the propagator and $F_{\pi}$ is the physical decay constant. In other words, they satisfy
\begin{align}
	& M_{\pi}^2 -M^2-\Sigma \big( M_{\pi}^2\big) = 0 \, , 
	\nonumber \\
	& \left. \langle 0 \right| A_{\mu} (0) \left| \pi (p) \rangle \right. = i F_{\pi} \sqrt{2}\, p_{\mu}  \, ,
\end{align}
where $\Sigma$ is the two-point function consisting of all amputated one-particle-irreducible diagrams, and $A_{\mu}$ is the axial current. From these relations it is possible to calculate $M_{2n}^2$ and $F_{2n}$, and the results take the forms of polynomials in chiral logarithms according to
\begin{align}
	& M_4^{ 2} = x\left(a_{10}^{ M}+a_{11}^{ M} L_M\right)
	\, , 
	\nonumber\\
	& M_6^{ 2} = x^2\left(a_{20}^{ M}+a_{21}^{ M} L_M+a_{22}^{ M} L_M^2\right)
	\, , 
	\nonumber\\
	& M_8^{ 2} = x^3\left(a_{30}^{ M}+a_{31}^{ M} L_M+a_{32}^{ M} L_M^2+a_{33}^{ M} L_M^3\right)
        \, ,
	\nonumber \\
	& F_4 = x\left(a_{10}^{ F}+a_{11}^{ F} L_M\right)
\, , 
	\nonumber\\
	& F_6 = x^2\left(a_{20}^{ F}+a_{21}^{ F} L_M+a_{22}^{ F} L_M^2\right)
\, , 
	\nonumber\\
	& F_8 = x^3\left(a_{30}^{ F}+a_{31}^{ F} L_M+a_{32}^{ F} L_M^2+a_{33}^{ F} L_M^3\right)
	\, .
\end{align}
Here the chiral logarithm is $L_{M} = \log M^2/\mu ^2$, $x = M^2/\big( 16 \pi ^2 F^2\big) $ and $a_{ij}^{F,M}$ are constants depending on the LECs. The leading logarithmic coefficients $a_{ii}^{M,F}$ can be determined without knowing the higher order Lagrangians and are known up to six-loop order~\cite{Bijnens:2013yca}. For massless $O(N)$ and $SU(N)$ models, the leading logarithms are known to any order~\cite{Ananthanarayan:2018kly}. Since the NNNLO Lagrangian only contributes at tree level, the NNNLO LECs cannot multiply a chiral logarithm. Therefore, they only appear in $a_{30}^{M,F}$ as two renormalised linear combinations $r_{M8}^{r}$ and $r_{F8}^{r}$. As was remarked in the previous section, these combinations have no known numerical values. However, at least from Table~\ref{table:nnnlonumbers} it can be deduced that at most 122 LECs enter into the combinations. For the $a_{3i}^{M,F}$ also NLO and NNLO LECs appear. The $\ell _{i}^{r}$ have known values, but not all the $c_{i}^{r}$. It is possible to rewrite some of the $c_{i}^{r}$ in terms of specific combinations appearing in $\pi \pi$ scattering, whose numerical values were estimated in Ref.~\cite{Bijnens:1997vq}. The unknown $c_{i}^{r}$ needed for the mass and decay constant at NNNLO are
\begin{enumerate}
	\item Mass: $c_{6}^{r}$,
	\item Decay constant: $c_{3}^{r}$, $c_{5}^{r}$, $c_{6}^{r}$, $c_{12}^{r}$, $c_{14}^{r}$, $c_{20}^{r}$.
\end{enumerate}
Despite not knowing these LECs, the authors of Ref.~\cite{Bijnens:2017wba} performed a small numerical study in terms of the quark mass dependence through $M^2 = 2B\hat{m}$, where $\hat{m} $ is the isospin symmetric quark mass. All unknown parameters were put to zero. The result is shown in Fig.~\ref{fig:mqdep}. There one can see how the the different orders in the chiral expansion compare. Around the physical point $M^2 \approx 0.02 \, \textrm{GeV}^2$ the contributions are in agreement, as they should. For the decay constant one sees that the NNNLO curve deviates drastically from the other two for higher quark masses. This comes from a very large $a_{30}^{F}\approx -244.5$, which in turn is an artefact of a large term from the loop calculation, $-383293667/1555200\approx -246.5$. Whether or not this large number persists once the unkown $c_{i}^{r}$ are known is impossible to say, but for them to give $a_{30}^{F}$ of a natural size, say $\sim1-10$, one would need $4c_{12}^{r}-2 c_{6}^{r}-c_{5}^{r}\approx A\times (16 \pi ^2)^{-2}\times 246.5 \approx A\times 0.01 $, where $A\sim 1-10$, which indeed is not impossible.

As a final remark, the analytic formulae for the pion mass and decay constant may of course be used for fits to lattice data in the future. In such a study it can be investigated how sizeable the NNNLO corrections are.

\section{Conclusions}
Chiral perturbation theory has been very successful in describing the low energy sector of QCD in terms of mesonic degrees of freedom. This effective field theory is expanded systematically order by order in powers of the momentum $p^2$. From the perspective of higher order calculations, the state-of-the-art currently is at NNNLO, or order $p^8$. The mesonic chiral Lagrangian is known, but renormalisation still needs to be done and numerical values for the LECs must be estimated for predictions from NNNLO. Also, no anomalous NNNLO chiral Lagrangian is currently known. The pion mass and decay constant are known for the isospin symmetric two-flavour case, but for these contributions also unkown NNLO LECs show up. One therefore needs to continue improving the knowledge of LECs at NNLO as well. For the moment, however, it is still possible to make continued analytic progress at NNNLO. 

\bibliography{refs}
\bibliographystyle{JHEPmy}

\end{document}